\documentclass[12pt,preprint]{aastex}

\slugcomment{Accepted by The Astrophysical Journal}

\shorttitle{Molecular environment of B1-IRS}
\shortauthors{de Gregorio-Monsalvo et al.}

\begin{document}

\title{High-Resolution Molecular Line Observations of the Environment of
  the Class 0 Source B1-IRS}

\author{Itziar de Gregorio-Monsalvo \altaffilmark{1,2}, Claire
  J. Chandler\altaffilmark{2}, Jos\'e F. G\'omez\altaffilmark{1},
  Thomas B. H. Kuiper\altaffilmark{3}, Jos\'e
  M. Torrelles\altaffilmark{4}, Guillem
  Anglada\altaffilmark{5}} 

\altaffiltext{1}{Laboratorio de Astrof\'{\i}sica Espacial y F\'{\i}sica
  Fundamental (INTA), Apartado 50727, E-28080 Madrid, Spain}
\altaffiltext{2}{National Radio Astronomy Observatory, P.O. Box O,
  Socorro, NM 87801, USA}
\altaffiltext{3}{ Jet Propulsion Laboratory, California Institute of Technology, USA}
\altaffiltext{4}{Instituto de Ciencias del Espacio (CSIC) and Institut d'Estudis Espacials de Catalunya, C/Gran Capit\`a 2-4, 
E-08034 Barcelona, Spain. On sabbatical leave at the UK Astronomy
Technology Center, Royal Observatory Edinburgh.}
\altaffiltext{5}{Instituto de Astrof\'{\i}sica de Andaluc\'{\i}a
  (CSIC), Apartado 3004, E-18080 Granada, Spain}

\begin{abstract}

In this work we present VLA observations of the NH$_3$,
CCS, and H$_2$O maser emission at 1 cm from the star forming region
\objectname{B1-IRS} (\objectname{IRAS 03301+3057}) with $\simeq 5"$ (=1750 AU) of angular resolution. CCS emission is distributed in three clumps around
the central source. These clumps exhibit a velocity gradient from red- to blueshifted velocities toward B1-IRS,
probably due to an interaction with the outflow from an embedded
protostar. The outflow and its powering source are traced by a
reflection nebula and an associated infrared point source detected in a 2MASS
K-band image. We find that this infrared point source is associated
with water maser emission distributed in an elongated structure 
($\simeq$ 450 AU size) along the major axis of the reflection nebula 
and tracing the base of the outflow of the region. 
Ammonia emission is extended and spatially anticorrelated
with CCS. This is the first time that this kind of anticorrelation 
is observed in a star forming region with such a high angular resolution, and illustrates the importance of time-dependent chemistry on
small spatial scales. 
The relatively large abundance of CCS with respect to ammonia, compared with
other star forming regions, suggests an extreme youth for the B1-IRS
object ($\leq$ 10$^5$ yr).
We suggest the possibility that CCS abundance is enhanced via
shock-induced chemistry. 
%However, theoretical calculations will be 
%needed to confirm this hypothesis.

\end{abstract}


\section{Introduction}

The spatial distribution of emission from various molecular tracers in
star forming clouds depends not only on the physical conditions within
the clouds but also on the (time-dependent) chemistry.  This is very
clearly the case for CCS and NH$_{3}$, where a pronounced spatial 
anticorrelation in the emission from these two species has been reported for
quiescent, starless cores \citep{Suz92}.  A spatial anticorrelation
has also been observed on large scale maps in several individual sources,
such as \objectname{TMC-1} \citep{Hir92}, \objectname{L1498} (Kuiper, Langer, \& Velusamy 1996), and \objectname{B68} \citep{Lai03}.

The lines of CCS are very appropriate for studying the structure and the
physical conditions in dark clouds because they are not very opaque
\citep{Sai87}, and yet they are intense and abundant
\citep{Suz92}.  Its lack of hyperfine splitting and the fact that it is
heavier than other high-density tracers (thus showing intrinsically
narrower lines) make CCS a very well-suited molecule for carrying out 
dynamical studies.  NH$_{3}$
emission tends to trace the inner, denser regions of molecular cores,
while CCS shows up slightly outside these central regions
with a clumpy distribution \citep{Suz92}. This relative distribution
has been explained
in terms of chemical and dynamical evolution of the protostellar core:
CCS lines are intense in cold quiescent cores, but when the cloud evolves
and star formation begins, high velocity outflows and radiation from
protostars destroy carbon chain molecules.  At the same
time, these conditions favor the desorption of NH$_3$ from dust grains.
For this reason, the abundance ratio [CCS]/[NH$_{3}$] has been considered
as an indicator of evolution in protostellar cores \citep{Suz92}.

CCS is rarely found to be associated with evolved star-forming regions.
On the other hand, emission from both CCS and NH$_{3}$ has been detected from
the Class 0 protostar \objectname{B335} (\citealt{Men83}; Velusamy, Kuiper, \&
Langer 1995) suggesting that a combination of both molecules can be
used as a means of identifying the youngest protostars, and of studying
their environments over a range of physical and chemical regimes.  Class 0
protostars have been proposed to be in the earliest stages of low-mass
stellar evolution (Andr{\' e}, Ward-Thompson, \& Barsony 1993).  Processes related to mass-loss phenomena
occur prominently during this phase: collimated jets, powerful molecular
outflows, and protoplanetary disks of hundreds of AU are commonly detected
in this kind of source \citep{Cha93, Ter93}.  CCS and NH$_{3}$ may
therefore trace the interaction of these phenomena with the surrounding
environment, providing information about the kinematics, temperature,
and density in the close vicinity of very young stellar objects (YSOs).

Water maser emission has also been shown to be a good tracer of mass-loss
activity in YSOs \citep{Rod80,Deb05}.  Previous studies suggest that water maser
emission provides a good characterization of the age of low-mass YSOs, with
Class 0 sources being the most probable candidates to harbor water masers
\citep{Fur01}. Water masers have been detected in both circumstellar
disks and outflows. This dichotomy has also been suggested to be a
time-dependent effect, with masers tracing disks in younger objects and
tracing outflows in somewhat more evolved sources \citep{Tor02}.

The far-infrared source B1-IRS (IRAS 03301+3057) is one of the few sources known to exhibit
both CCS and water maser emission in the studies by \citet{Suz92} and
\citet{Fur01}. This is a Class 0 source \citep{Hir97}
located in the Perseus OB2 complex at a distance of 350 pc (Bachiller, Menten, \& del Rio {\' A}lvarez 1990).
The \objectname{B1} cloud contains a large amount of molecular gas, as shown by the
strong C$^{18}$O(1--0) emission detected by \citet{Bac84}.  There is no
optically visible counterpart of B1-IRS, but 850 micron dust continuum
emission has been detected by \citet{Mat02} using SCUBA\@.  There are
two compact SiO clumps in the vicinity of B1-IRS \citep{Yam92}, and
blueshifted CO(1--0) emission observed by \citet{Hir97}, who suggested
the presence of a pole-on molecular outflow.  The SiO clumps are
located at the interface between the CO outflow and the dense gas traced
by the C$^{18}$O, suggesting that the outflow, probably driven by the
IRAS source, is interacting with the surrounding medium, thus yielding
the SiO emission through shocks. 

In this work, we have used the Very Large Array (VLA) of the National
Radio Astronomy Observatory\footnote{The NRAO is operated by Associated
Universities Inc., under cooperative agreement with the National Science
Foundation.} to study the properties of the CCS, NH$_{3}$, and water
maser emission from B1-IRS, in order to obtain information about its
circumstellar dynamics and stage of evolution.  The rest frequencies
of the CCS and H$_2$O transitions are sufficiently close together
that they can be observed simultaneously with the VLA, also offering
the possibility of using the water masers to track tropospheric phase
fluctuations, if the masers are strong enough.  Unfortunately this was
not the case at the time of the observations presented here, but the
atmosphere was in any case very stable.

This paper is structured as follows: in \S 2 we describe our observations
and data reduction procedure, as well as the analysis of VLA archive
data. In \S 3
we present our observational results (morphology and physical parameters),
which are then discussed in \S 4.  We summarize our conclusions in \S 5.

\section{Observations and data processing}

Simultaneous observations of the J$_{N}$=2$_{1}$-1$_{0}$ transition 
of CCS (rest frequency $=$ 22344.033 MHz) and the 6$_{16}$-5$_{23}$ transition
of H$_{2}$O (rest frequency = 22235.080 MHz) were carried out on 2003
April 4 using the VLA in its D configuration.  The phase center of these
observations was R.A.(J2000) = 03$^{h}$33$^{m}$16$\fs$3, Dec(J2000) =
31$\degr$07$\arcmin$51$\arcsec$.  We used the four IF spectral line mode,
with two IFs for each molecular transition, one in each of right and left
circular polarizations.  For the CCS observations we sampled 128 channels
over a bandwidth of 0.781 MHz centered at $V_{\rm LSR} =  6.8$ km~s$^{-1}$, 
with a velocity resolution of 0.082
km~s$^{-1}$.  Water maser observations were obtained using 64 channels
over a 3.125 MHz bandwidth centered at $V_{\rm LSR} = 15.7$ km~s$^{-1}$, with 0.66 km~s$^{-1}$ velocity resolution.
The total on-source integration time was $\simeq$ 7.5 hours. Our primary
calibrator was 3C48, for which we adopted a flux density of 1.1253 Jy
(at the
CCS frequency) and 1.1315 Jy (at the H$_{2}$O frequency) using the latest
VLA values (1999.2). The source 3C84 was used as phase and bandpass
calibrator (bootstrapped flux density = 11.80$\pm$0.14 Jy at the CCS
frequency and 11.65$\pm$0.14 Jy at the H$_{2}$O frequency). Calibration
and data reduction were performed with the Astronomical Image Processing
System (AIPS) of NRAO.

CCS data were reduced applying spectral Hanning smoothing and natural
weighting of the visibilities, to improve the signal-to-noise ratio.
The final velocity resolution is 0.16 km s$^{-1}$ and the synthesized beam
is 4$\farcs$33 $\times$ 3$\farcs$30 (P.A = $-$79$\fdg$2).  Water maser
maps were produced using a ``robust'' weighting \citep{Bri95} of 0, to optimize a
combination of both angular resolution and sensitivity. The synthesized
beam obtained is 3$\farcs$29 $\times$ 2$\farcs$73 (P.A = 87$\fdg$9).

In order to compare our water maser results with those at other epochs,
we also retrieved earlier H$_{2}$O data from the VLA archive. These
observations were carried out on 1998 October 24 and 1999 February 26,
in CnB and CD configuration respectively, for project AF354. These data
have been published by \citet{Fur03}. Both sets of observations were
made in the 1IF spectral line mode, in right circular polarization
only, with a bandwidth of 3.125 MHz centered at $V_{\rm LSR} = 0$ km~s$^{-1}$,
and sampled over 128 channels, thus
providing a velocity resolution of 0.33 km~s$^{-1}$. The phase center
of these observations was R.A.(J2000) = 03$^{h}$33$^{m}$16$\fs$030,
Dec(J2000) = 31$\degr$07$\arcmin$34$\farcs$05.  The time on source was
$\simeq$ 3 minutes in 1998 and $\simeq$ 13 minutes in 1999. In both epochs 3C48
was used as primary flux calibrator, with an assumed flux density of
1.1313 Jy and 1.1315 Jy for the 1998 and 1999 observations respectively
(adopting the latest 1999.2 VLA values). The secondary calibrator was
0333+321, with a bootstrapped flux density of 1.66$\pm$0.10 Jy for
the 1998 observations and 1.67$\pm$0.03 Jy for the 1999 observations.
The water masers were strong enough during these earlier observations
to enable self-calibration, after referencing the maser positions
to 0333+321.  Images were made with the robust parameter set to 0, giving
synthesized beams of $0\farcs91 \times 0\farcs38$ (P.A. = 81$\fdg$7)
and $3\farcs05 \times 1\farcs16$ (P.A. = 68$\fdg$8) for the 1998
and 1999 data respectively.

We have also processed VLA archive data of the NH$_{3}$(1,1) inversion
transition (rest frequency = 23694.496 MHz). The observations
were made on 1988 August 13 for project AG265, in the D
configuration. The total
bandwidth was 3.125 MHz, centered at $V_{\rm LSR} = 6.1$  km~s$^{-1}$ and 
sampled by 128 channels, which provided
a velocity resolution of 0.33 km~s$^{-1}$. Both right and left
circular polarizations were obtained. The phase center of these
observations was R.A.(J2000) = 03$^{h}$33$^{m}$16$\fs$337, Dec(J2000) =
31$\degr$07$\arcmin$51$\farcs$03. The total time on source was
$\simeq$ 2 h.
The source 3C48 was used as the primary flux calibrator, with an adopted
flux density of 1.0542 Jy (1999.2 values). The phase calibrator was 3C84
(bootstrapped flux density = 38.0$\pm$1.5 Jy).  Since the NH$_{3}$ emission
was faint and extended, we reduced the data using natural weights and
a {\it uv}-taper of 20 k${\lambda}$, to improve the sensitivity to the
extended emission.  The resulting synthesized beam is $8\farcs62 \times
7\farcs76$ (P.A. = $-75\fdg5$).
The size of the VLA primary beam at the frequencies corresponding to
the CCS, H$_{2}$O and NH$_{3}$ transitions ($\simeq$22 GHz) is $\simeq$ 2$\arcmin$.

To have an estimate of the missing flux density due to the lack of short spacings of
the VLA, we have obtained single-dish spectra of both CCS and NH$_3$
towards the position  R.A.(J2000) = 03$^{h}$33$^{m}$16$\fs$3, Dec(J2000) = 
31$\degr$07$\arcmin$51$\arcsec$,
with the NASA's 70 m antenna (DSS-63) at Robledo de Chavela, Spain, at
the same frequencies as the VLA. This antenna has a 1.3 cm
receiver comprising a cooled high-electron-mobility
transistor (HEMT) amplifier. A noise diode is used to calibrate the data. 
The half power beam width at this frequency is $\simeq$ 41$''$, and the
mean beam efficiency is $\simeq$ 0.4. Observations were made in frequency
switching mode, using a 256
channel autocorrelator spectrometer. The
CCS observations were performed in six time slots between 2002 August and  2003 July with a total
integration time of 60 minutes and an average system temperature of
$\simeq$ 80 K. We
used a bandwidth of 1 MHz (velocity resolution
= 0.05 km s$^{-1}$).  
The NH$_3$ observations were carried out during 2003 July, with an 
average system
temperature of $\simeq$ 60 K. The on-source integration time was 40 minutes. 
To detect all the hyperfine lines we used a bandwidth of 10 MHz 
(velocity resolution = 0.5 km s$^{-1}$). The rms pointing accuracy of
the telescope was
better than 9$''$. 
All the single-dish data reduction was carried out using the CLASS package,
developed at IRAM and the Observatoire de Grenoble as part of the GAG
software.

In addition to the radio data, we have retrieved a K-band image from
the Two Micron All Sky Survey (2MASS), to
obtain a better position for the infrared source. This image was smoothed
with a 5$\arcsec$ Gaussian (FWHM), to search for low surface-brightness,
extended emission.

\section{Results}

\subsection{The infrared source}

\begin{figure} 
\rotatebox{-90}{
\plotone{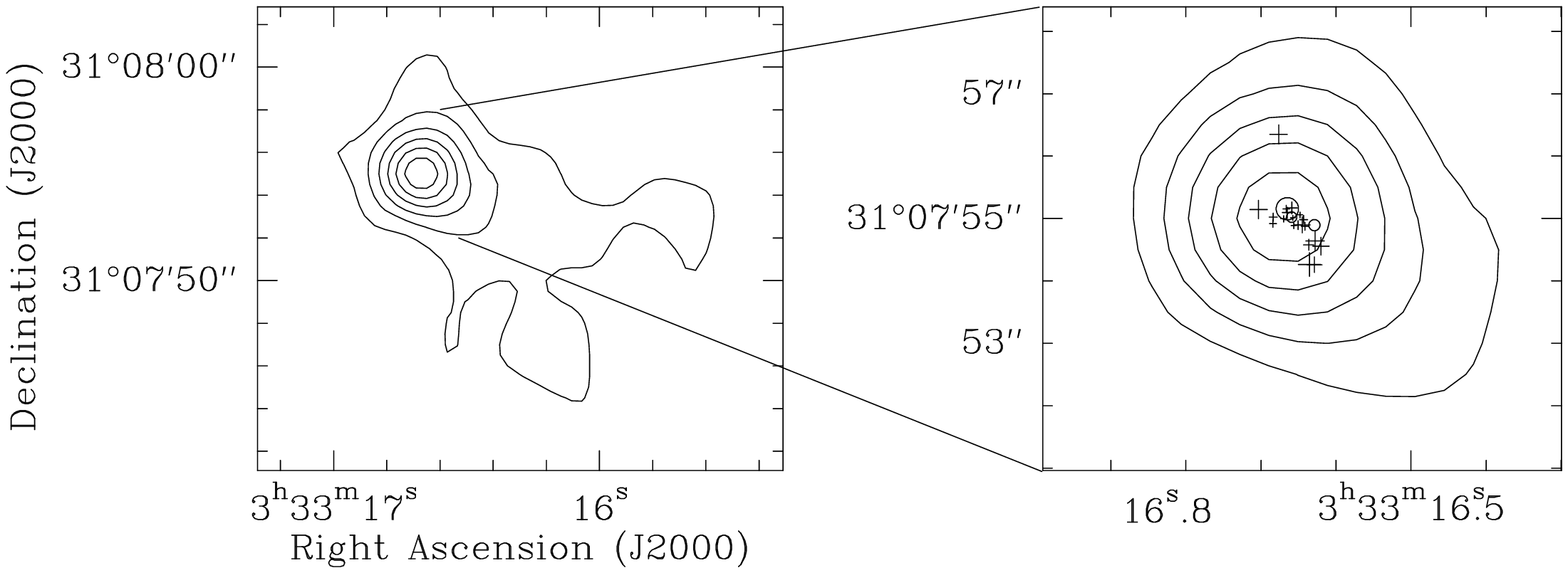}} 
\caption{2MASS K-band image convolved with a 5$\arcsec$
  Gaussian. Contours represent 90\%, 75\%, 60\%, 45\%, 30\%, and  15\%
  of the peak emission (the source magnitude is K = 14.208$\pm$0.122, measured
  by 2MASS). Crosses and open circles in the close-up of the infrared emission (right panel) represent the structure of red and blueshifted
 water masers respectivelly, obtained in the April 2003
  observations and scaled by their relative position uncertainty.}\label{2MASS} 
%The upper panel represents the original 2MASS
%  K-band image (see the arrow) in grey scale. The lower pannel represents the
%2MASS K-band image convolved with a 5$\arcsec$ Gaussian.
% Contours represent 15\%, 30\%, 45\%, 60\%, 75\%, and 90\%
%  of the peak emission (the source magnitude is K = 14.208$\pm$0.122, measured
%  by 2MASS). Crosses and open circles in the close-up of the infrared emission (right panel) represent the structure of red and blueshifted
%  water masers respectivelly, obtained in the April 2003
 % observations and scaled by their relative position uncertainty.
\end{figure}

The 2-$\micron$ map obtained from the 2MASS archive, and convolved with
a 5$\arcsec$ Gaussian (Fig.~\ref{2MASS}), shows a point source,
designated \objectname{2MASS J03331667+3107548}, at R.A.(J2000) = 03$^{h}$33$^{m}$16$\fs$678, Dec(J2000)
= 31$\degr$07$\arcmin$54$\farcs$88 (2$\sigma$ absolute position error $\simeq
0\farcs$22).  This position is $\simeq 6\arcsec$ away from the IRAS
catalog position, R.A.(J2000) = 
03$^{h}$33$^{m}$16$\fs$3, Dec(J2000) = 31$\degr$07$\arcmin$51$\arcsec$,
but well within the error ellipsoid (2$\sigma$ error axes =
48$\arcsec$$\times$14$\arcsec$) of the latter.  In what follows, we
will consider the position of the 2MASS point source as the location
of the central source of B1-IRS. 
We also find extended 2 $\mu$m infrared emission elongated to the south
west of the point source, with position angle $\simeq -120\degr$,
suggestive of a reflection nebula.

\subsection{The water masers}

Our water maser observation on April 2003 shows a cluster of 23
 maser spots (Table~\ref{tbl-april}), spatially unresolved in each
 channel, most of which are redshifted with respect to
the cloud velocity ($V_{\rm LSR} = 6.3$ km s$^{-1}$;
\citealt{Hir97}) with LSR velocities between 9.1 and
23.6 km s$^{-1}$, although three are blueshifted, with LSR velocities between
2.5 and 3.8 km s$^{-1}$.  The maximum flux density was $\simeq$ 15.4$\pm$1.6
mJy at $V_{\rm LSR}$ = 17.7 km~s$^{-1}$,
two orders of magnitude weaker than in the
earlier epochs reported by \citet{Fur03}.  The masers are distributed
in an elongated structure (P.A. $\simeq -120 \degr$), with a length of
$\simeq$ $1$\farcs$3$ ($\simeq$455 AU).

\begin{figure}
\rotatebox{-90}{
\epsscale{0.6}
\plotone{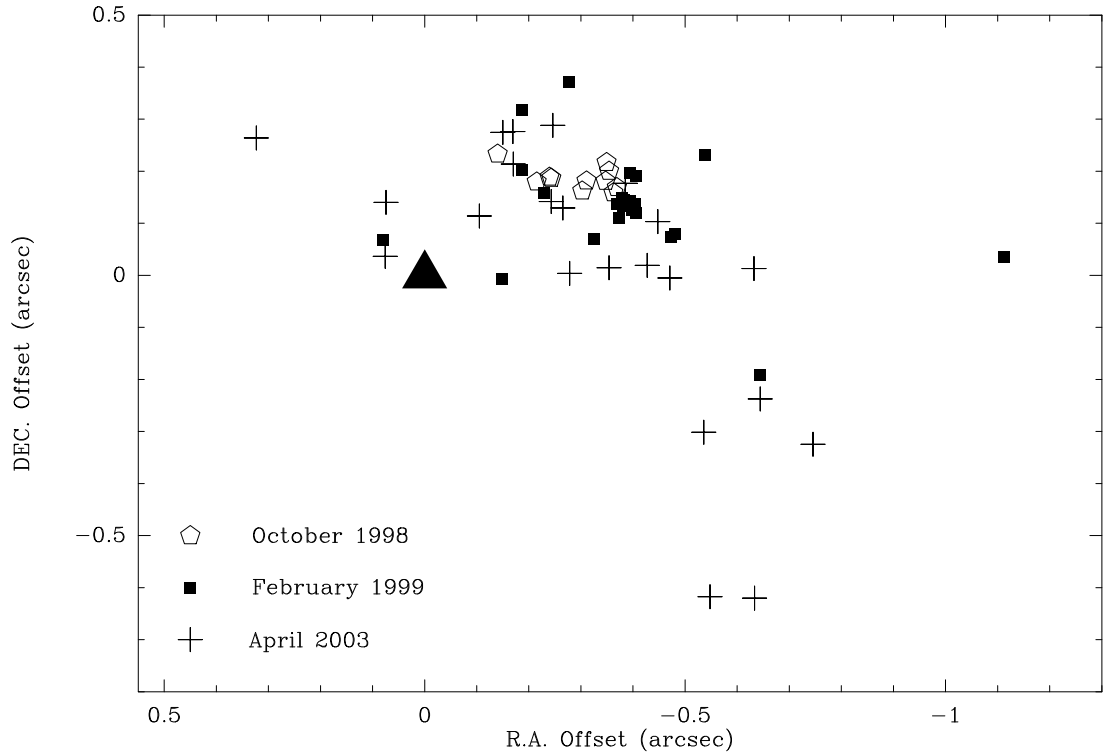}}
\caption{Positions of water masers observed towards B1-IRS at
  different epochs. 
Open pentagons, filled squares, and crosses represent the maser components
observed in  October 1998,  February  1999, and April 2003, respectively. 
The position offsets are relative to the 2 micron point source, which is
represented by a filled triangle. The northernmost maser spot shown in
Fig. \ref{2MASS} is out of the scale in this plot. \label{masers}} 
\end{figure}

We have compared our new data with those obtained in 1998 October and
1999 February by \citet{Fur03}. The maser positions for those epochs are
shown in Fig.~\ref{masers}, and in Tables~\ref{tbl-oct} and \ref{tbl-feb}. The
same elongated distribution is evident. In these
earlier data sets all maser emission is redshifted with respect to the
cloud velocity.  We note that there is a discrepancy of $\simeq$23$\arcsec$
between the positions of the 1998 October masers presented here and the
positions reported in Table 3 of \citet{Fur03} using the same data. The positions
of the masers (that we have determined by fitting two-dimensional Gaussians to the
maser spots detected in each channel), are similar in all three epochs,
taking into account their mean absolute positional uncertainties ($\simeq$ 0$\farcs$3,
0$\farcs$16, and 0$\farcs$04 for 2003 April, 1999 February and 1998
October respectively). Therefore we are confident that our positions
are correct.

The water masers are located at a distance $\la$ 1$\arcsec$ (350 AU)
from the 2$\micron$ source, suggesting that this is the center
of activity of the region.  Considering an average LSR velocity of
$\simeq$ 14.0 km s$^{-1}$ of the maser emission, i.e., $\simeq$ 7.7 km
s$^{-1}$ redshifted with
respect to the cloud velocity (6.3 km s$^{-1}$), and a mean
distance of $0\rlap.''5$ from the central source, the central mass needed for the water masers to be gravitationally bound is $\simeq 12$
M$_{\sun}$. However, B1-IRS is a low luminosity protostar ($L \simeq 2.8$
L$_{\sun}$, \citealt{Hir97}), and it is unlikely to be much more massive than
$\simeq 1$ M$_\odot$. These results strongly 
suggest that the masers are tracing mass loss motions rather than
bound motions in a circumstellar disk.

\subsection{CCS and NH$_3$ emission}

\subsubsection{CCS emission}

\label{sec-ccs}

\begin{figure}
\rotatebox{-90}{
\epsscale{0.6}
\plotone{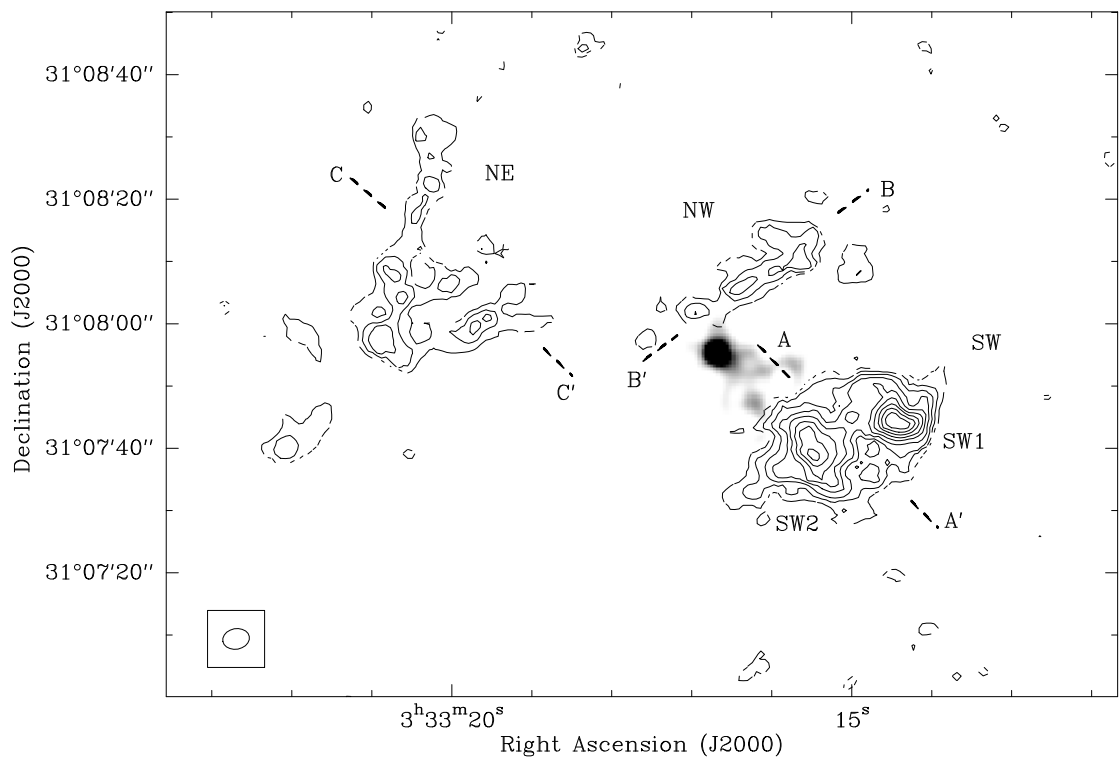}}
\caption{CCS integrated emission (contours) not corrected for the
  response of the VLA primary beam, overlaid on the 2MASS
  K-band emission (grey scale). Contour levels range from
  0.66 to 5.94 mJy beam$^{-1}$ km s$^{-1}$ in intervals of 0.66 mJy
  beam$^{-1}$ km s$^{-1}$. Grey scale ranges between 10\%
and 45\%  of the peak emission (the source magnitude is K =
14.208$\pm$0.122 measured by 2MASS). The south-west (SW), 
north-west (NW) and north-east (NE) CCS clumps are labeled.     
Dashed lines represent the axes where position-velocity diagrams have
been obtained (Fig. \ref{P-V}) .} \label{CCS-2MASS}

\end{figure}

\begin{figure}
\rotatebox{-90}{
\epsscale{0.9}
\plotone{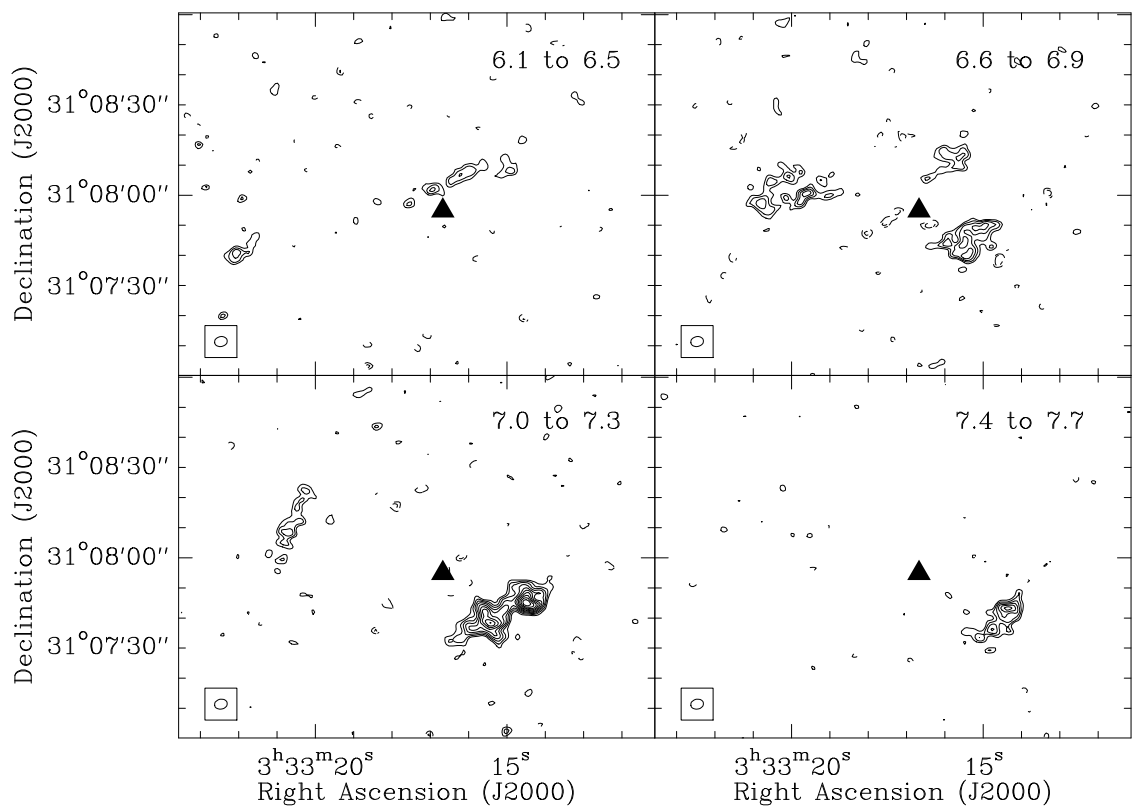}}
\caption{Contour maps of the CCS integrated emission over the LSR velocity ranges
  (in  km s$^{-1}$) indicated in the top right corner of each panel. Contour
  levels are at $-2.56$, $-1.92$, and from $1.92$ to $7.68$ at 
  steps of 0.64 mJy beam$^{-1}$ km s$^{-1}$ (the rms of the maps). 
  Filled triangle indicates the 2 $\mu$m point source position.
  No primary beam correction has been applied to these figures.}\label{CCSchannels}
\end{figure}

\begin{figure}
\rotatebox{-90}{
\epsscale{0.6}
\plotone{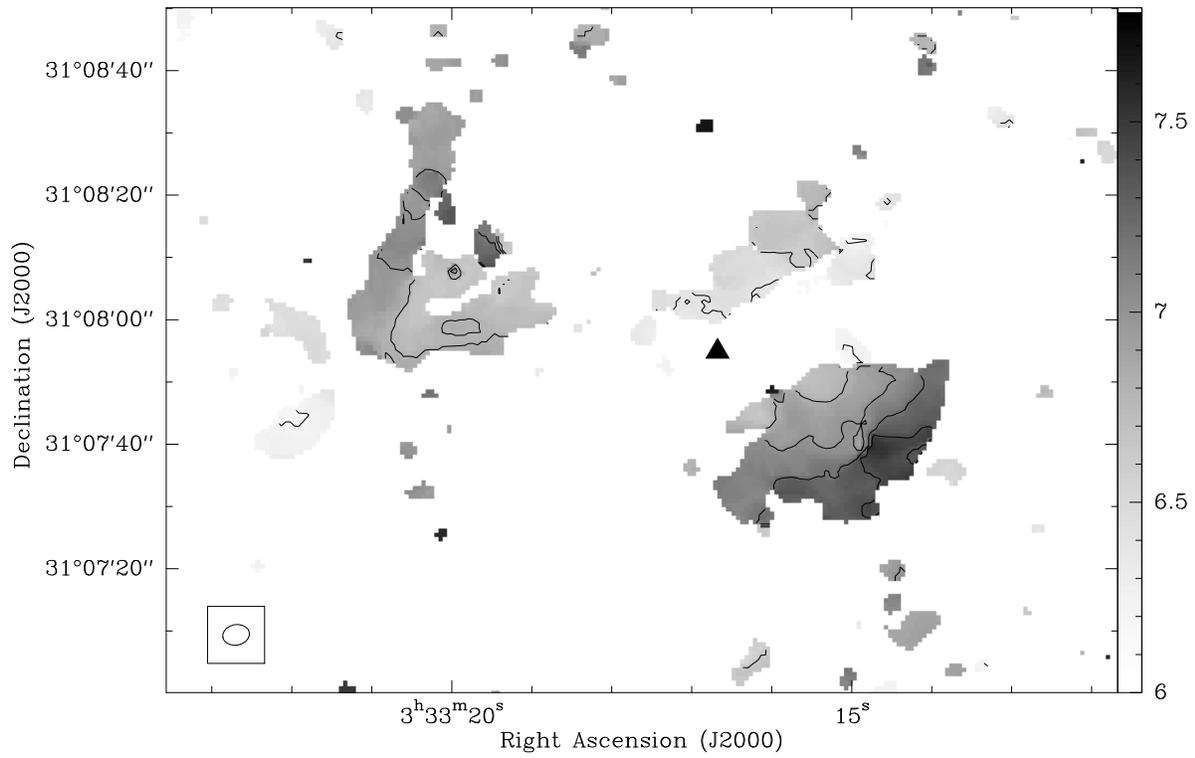}}
\caption{Map of CCS intensity-weighted mean velocity (first-order
  moment). Contour levels range from 6.0  to 7.8 km s$^{-1}$ at steps
  of 0.2 km s$^{-1}$. The grayscale is
  also from  6.0  to 7.8 km s$^{-1}$. The filled triangle represents the
  position of the 2  $\mu$m source.} \label{CCSvelocity}
\end{figure}

\begin{figure}
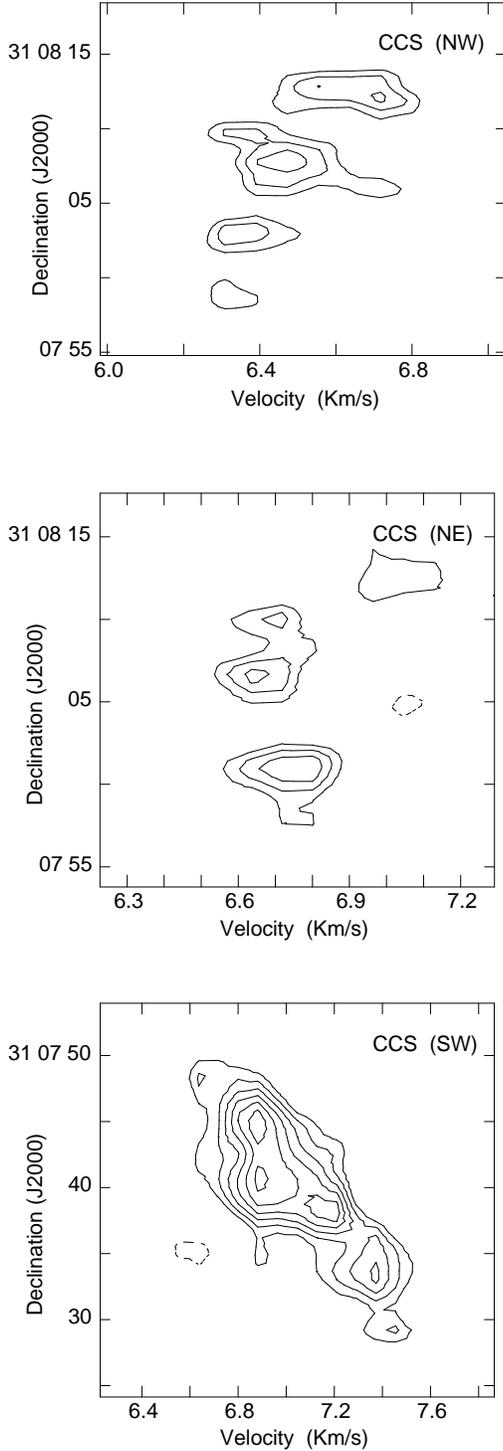


\rotatebox{-90}{
\epsscale{0.35}
\plotone{f6a.eps}}\\ \\
\rotatebox{-90}{
\epsscale{0.35}
\plotone{f6b.eps}}\\ \\
\rotatebox{-90}{
\epsscale{0.35}
\plotone{f6c.eps}}

\caption{Position-velocity diagrams of the CCS clumps. 
The upper panel represents the position-velocity diagram of the  NW
clump  through axis B-B' (see Fig. \ref{CCS-2MASS}), the central one corresponds to the NE clump
  through axis C-C', and the bottom  one the SW 
clump  through axis A-A'.
 Contours are at $-3.6$, and from $3.6$ to $9.6$ mJy beam$^{-1}$  with
 steps of 1.2 mJy beam$^{-1}$, the rms of the maps. These plots have
 not been corrected from the response of the VLA primary beam.\label{P-V}}
\end{figure}

The integrated emission from the CCS molecule is distributed in
three clumps (Fig.~\ref{CCS-2MASS}).  
 The clump centers are located $\simeq$ 25$\arcsec$ (9000 AU) south-west
(SW clump), $\simeq$ 50$\arcsec$ (17000 AU) north-east (NE clump),
 and $\simeq$ 15$\arcsec$ (5000 AU) north-west (NW clump) from the central source.
Fig.~\ref{CCSchannels} shows the CCS emission
integrated over different velocity intervals.  All clumps
are redshifted with respect to the systemic velocity of the B1 core
($V_{\rm LSR}$ = 6.3 km~s$^{-1}$). Moreover, the SW, NE and NW CCS
clumps show clear velocity gradients of $\simeq$ 23,
$\simeq$ 10, and $\simeq$ 12 km~s$^{-1}$~pc$^{-1}$ respectively 
(see Figs.~\ref{CCSvelocity} and \ref{P-V}) over the whole size of the
clumps, with velocities closer to that of the ambient cloud near the
central object.

 Clump SW has the most
redshifted velocities ($V_{\rm LSR}$ from 6.6 km s$ ^{-1}$ to 7.6 km s$^{-1}$). It
has an ellipsoidal shape, elongated perpendicularly to the direction
of the velocity gradient,
with two emission peaks.  Clump NE shows emission from $V_{\rm LSR}$ =
6.6 km~s$^{-1}$
to 7.1 km~s$^{-1}$, and it is divided into three ``fingers'' that point towards
the north-west. 
Clump NW has a velocity gradient from $V_{\rm LSR}$ = 6.3 to 6.8 km~s$^{-1}$ (i.e.,
closer to the mean cloud velocity), and its morphology is elongated,
with its major axis pointing towards the central source.

\subsubsection{NH$_{3}$ emission}

\begin{figure}
\rotatebox{-90}{
\epsscale{0.7}
\plotone{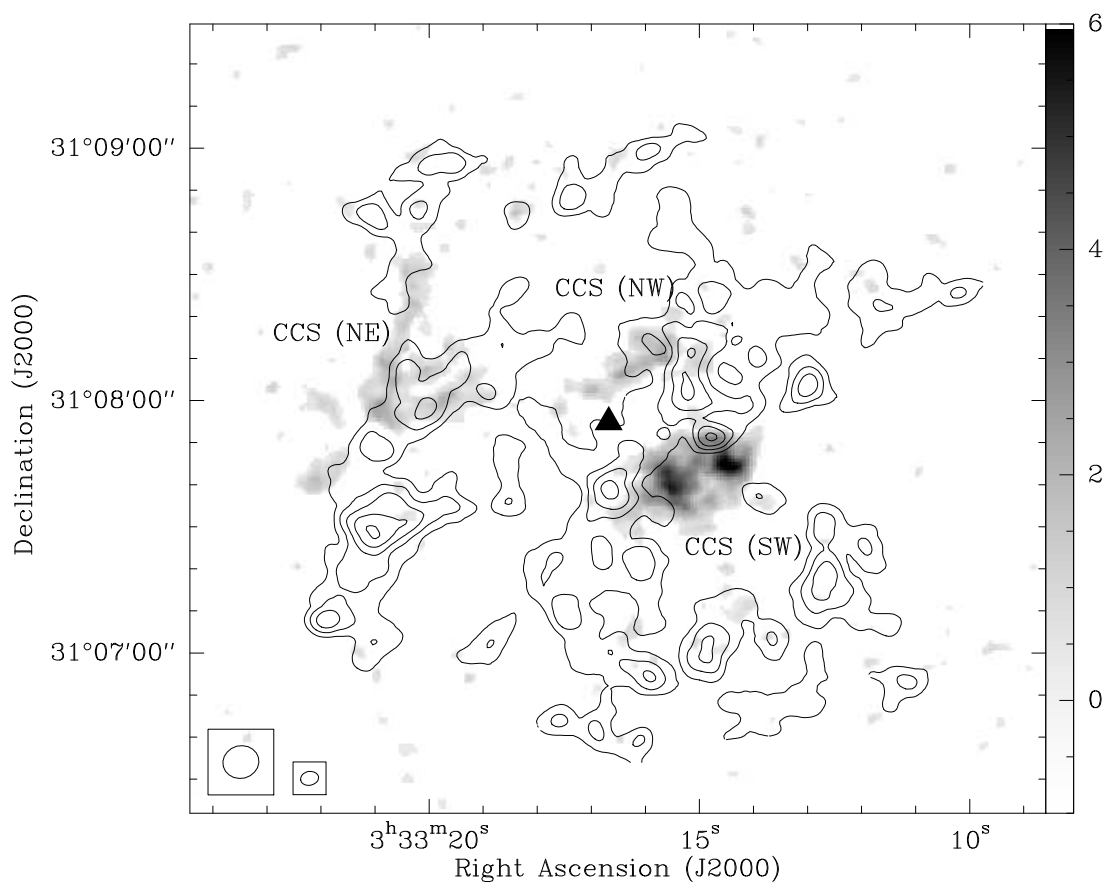}}
\caption{CCS (greyscale) and  ammonia main line (contours) integrated emission. Greyscale
 ranges from -1 to 6 mJy beam$^{-1}$ km
  s$^{-1}$. Contour levels range 
  from 40 to 120 mJy  beam$^{-1}$ km s$^{-1}$ with increment steps of 20  mJy
  beam$^{-1}$ km s$^{-1}$ , the rms of the map. Filled triangle
  represents the 2 $\mu$m source. The beam on the left correspond to
  the ammonia observations and the beam on the right corresponds to
  the CCS observations. No primary beam correction has been applied to
  these maps. However, given that the phase center was almost the same
  in the CCS and NH$_{3}$ observations, the beam response is similar
  and does not affect the comparison between both emissions.} 
  \label{CCS-NH3} 
\end{figure}

Ammonia emission is very extended and clumpy
(Fig.~\ref{CCS-NH3}).
Most clumps are distributed in two strips, with NW-SE
orientation.
The general trend, however, as illustrated by Fig.~\ref{CCS-NH3}, is that CCS and NH$_{3}$
emissions are spatially anticorrelated. Such an anticorrelation has been
observed in other sources \citep{Hir92, Wil98, Lai03}, but this is the
first time that it has been detected with such a high angular resolution ($\simeq$ 5$\arcsec$).

\begin{figure}
\rotatebox{-90}{
\epsscale{0.32}
\plotone{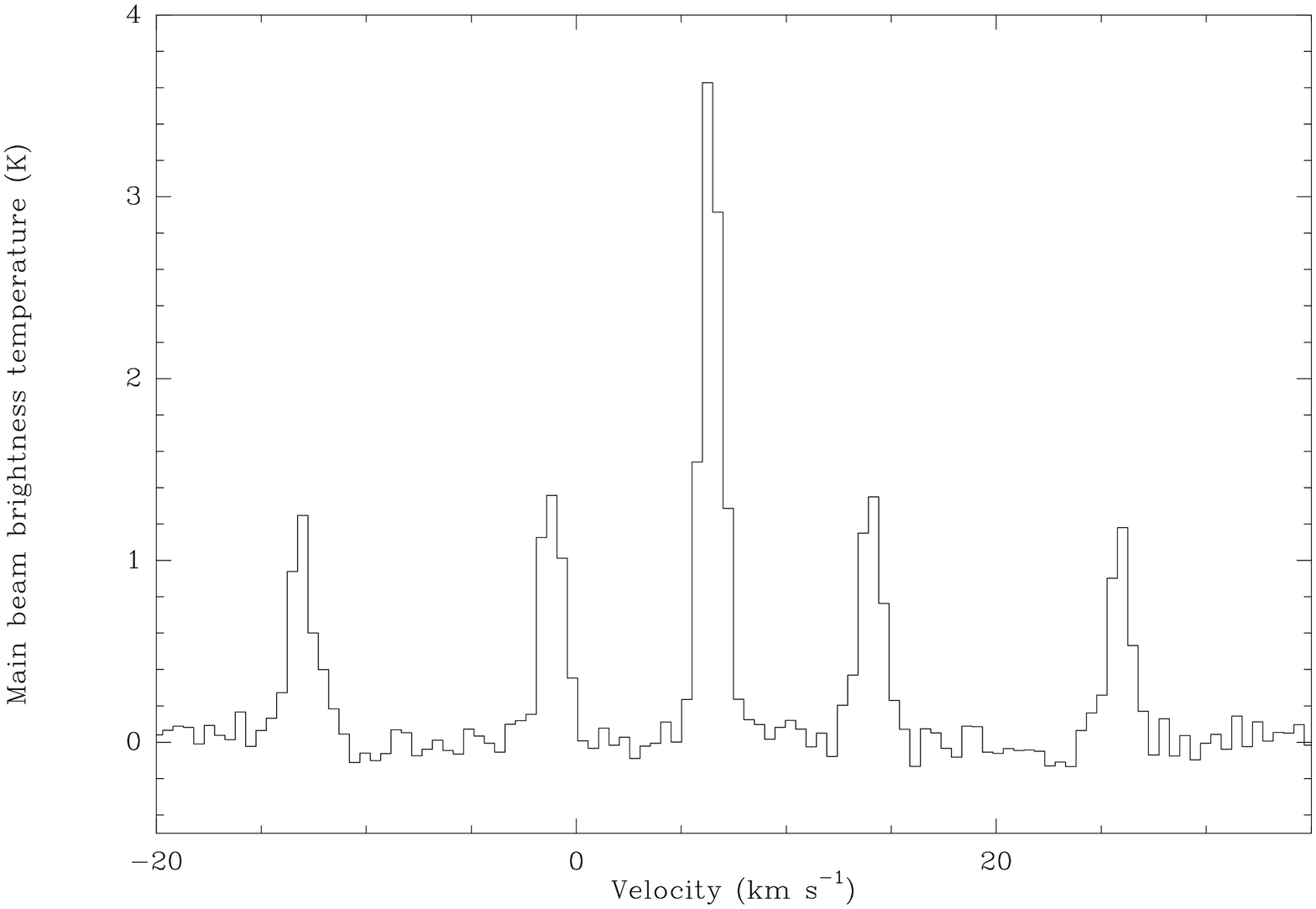}}
\rotatebox{-90}{
\epsscale{0.32}
\plotone{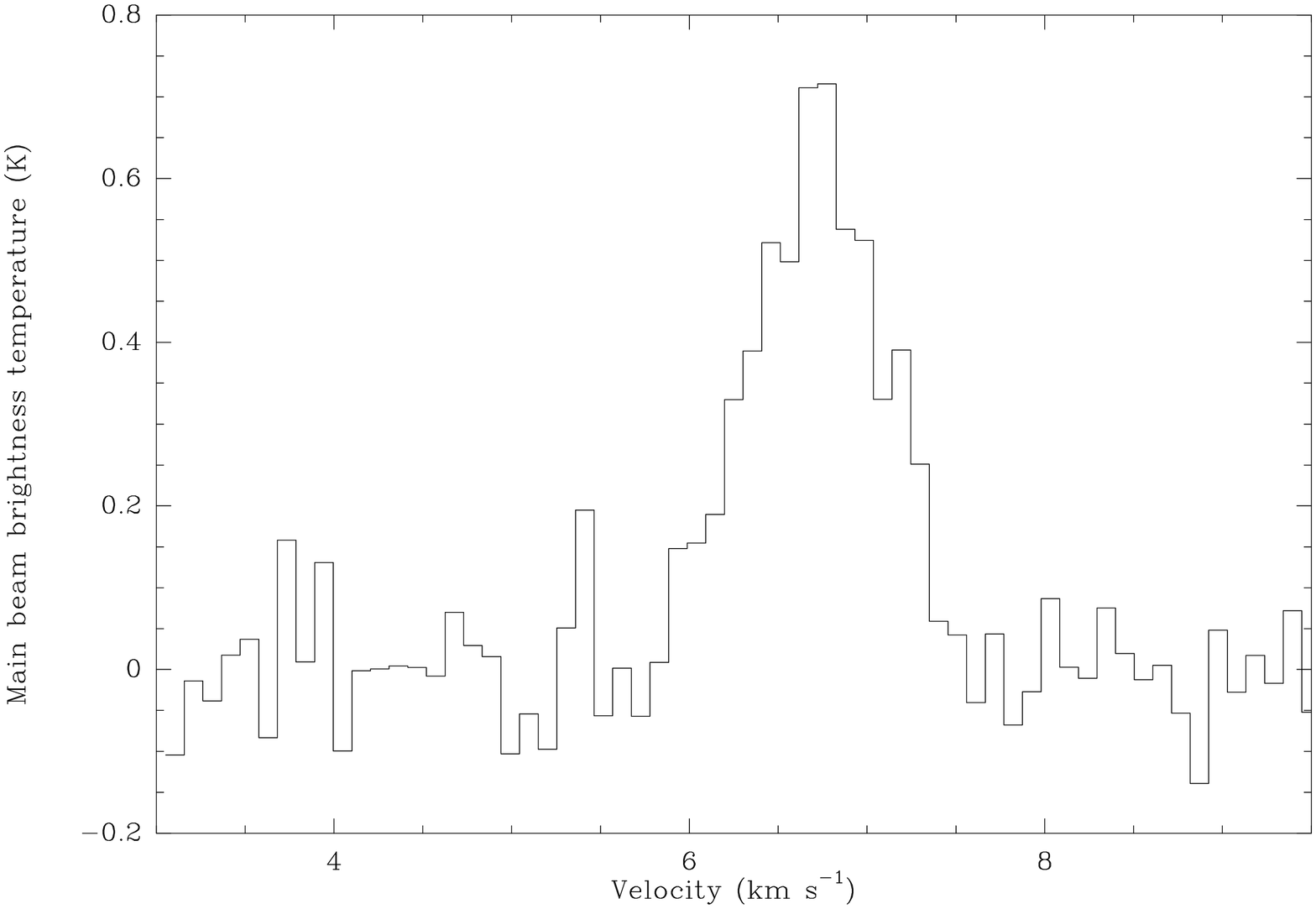}}
\caption{Single-dish spectra observed with the Robledo 70m antenna, at the
  position R.A.(J2000) = 03$^{h}$33$^{m}$16$\fs$34, Dec(J2000) = 
31$\degr$07$\arcmin$51$\farcs$1.
Left panel represents the NH$_{3}$(1,1) emission, with rms noise $1\sigma
=0.08$ 
K. Right panel is the CCS(2$_{1}$-1$_{0}$) spectrum, with rms noise  $1\sigma
=0.13$ K.}\label{Single_dish}
\end{figure}

We have to note, however, that there is considerable extended
NH$_{3}$(1,1) 
and  CCS (J$_{N}$=2$_{1}$-1$_{0}$) emission in the B1-IRS region
to which the VLA is not sensitive. Indeed, from our single-dish
observations of these transitions (Fig.~\ref{Single_dish}) we estimate
that $\simeq$ 90\% of the
ammonia and CCS emission is missed with the VLA.

\subsubsection{Physical parameters}

Although the maps shown in this paper have not been corrected by the primary
beam response for display proposes, such a correction has been applied
to the data prior to deriving physical parameters. Table~\ref{tbl-physical} summarizes the main physical
parameters obtained from the CCS and NH$_3$ lines.
For the CCS data, we provide parameters for
each individual clump, since they are well defined in the integrated intensity
map (Fig.~\ref{CCS-2MASS}). For these calculations, we have divided the
emission of the SW clump in two 
different peaks,  SW1 
(located at lower right ascension), and SW2 (located at the eastern part
of the SW clump). 
The mean clump column density of CCS is $N_{\rm CCS} \simeq (3.0\pm0.8)\times
10^{13}$cm$^{-2}$, which is similar to the value
obtained by \citet{Suz92} for B1 ($1.38\times 10^{13}$
cm$^{-2}$). Assuming a relative abundance of CCS with respect to H$_2$ of
$0.9\times10^{-10}$ (mean value reported by \citealt{Lai00} for B1), the
resulting mean $N_{\rm H_2}$ is  $\simeq (3.3\pm0.8) \times 10^{23}$
cm$^{-2}$. 

Given the low signal-to-noise
ratio and the clumpy structure of the NH$_3$ emission, we only give
averaged values for this molecule in Table~\ref{tbl-physical}. The
column density of ammonia at each local maximum 
seen in the integrated intensity map (Fig.~\ref{CCS-NH3}), ranges 
between $\simeq 6\times 10^{14}$ and $3\times 10^{15}$
cm$^{-2}$, with a mean clump value of $\simeq (1.4\pm0.4)\times 10^{15}$ cm$^{-2}$
, similar to the value of $1 \times 10^{15}$
cm$^{-2}$  reported by \citet{Bac86}.

\section{Discussion}

\subsection{The geometry of the molecular outflow}

\begin{figure}
\rotatebox{-90}{
\epsscale{0.7}
\plotone{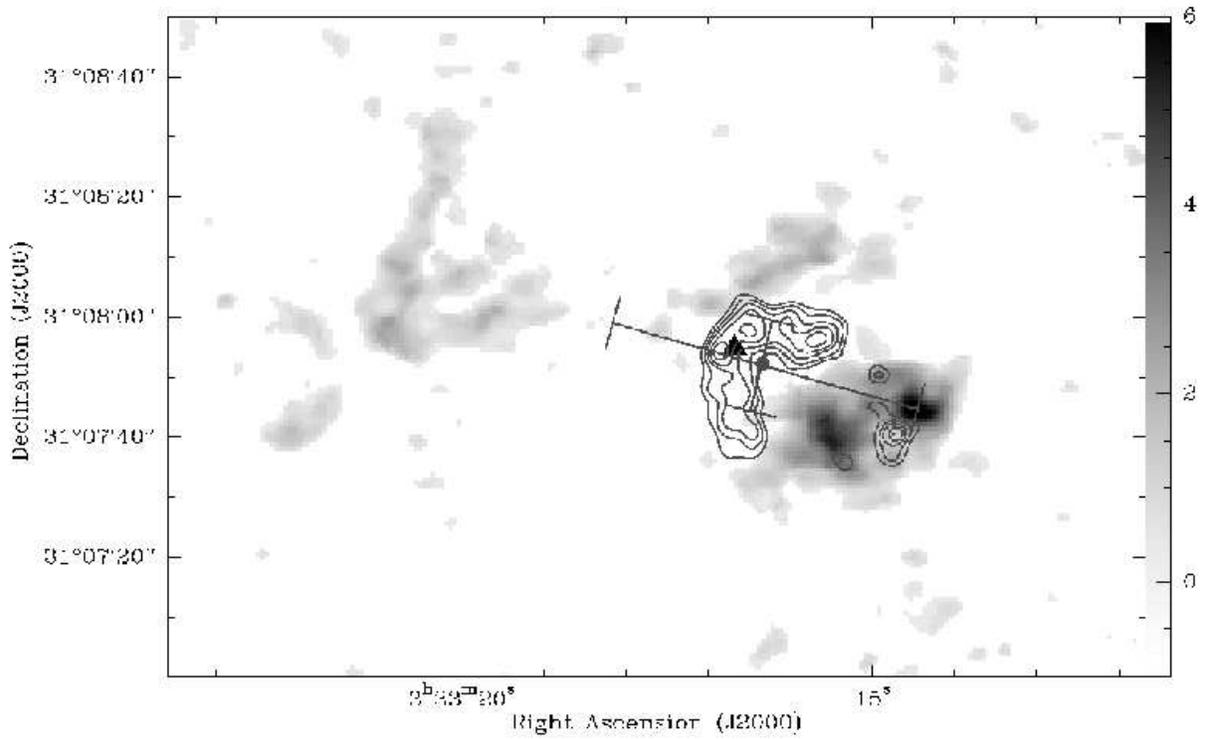}}
\caption{Overlay of the CO blueshifted outflow (contours; \citealt{Hir97})
  and CCS integrated emission (greyscale; this paper). The filled
  circle represents the IRAS source with its error position barrs. The
  triangle marks the 2 micron point source. The CCS emission is not
  corrected from the response of the primary beam.\label{CCS_CO}} 
\end{figure}

\citet{Hir97} detected moderately-high velocity blueshifted CO emission
centered close to the position of the IRAS source and consisting of
many clumps at different blueshifted velocities. No redshifted emission
was obvious in their data. 
Given the distribution of the blueshifted gas around the IRAS
position (Fig.~\ref{CCS_CO}), \citet{Hir97} suggested the existence of
a molecular
outflow driven by the IRAS source, and oriented pole-on with respect to the observer. 
In order to explain the absence of the 
redshifted CO lobe, these authors suggested that higher density 
material is slowing down that component of the outflow.

Our results, however, do not support the pole-on geometry for the
outflow. The 2MASS source, which is within the error box of the IRAS
position, is located at the tip of the blueshifted outflow, rather
than at its center (see Fig.~\ref{CCS_CO}). 
If we assume that the 2MASS source is actually tracing the powering source of the outflow
(confirmed by its association with the water masers), its blueshifted
lobe would flow towards the southwest from the central source. 

The low velocity  of the CO  blueshifted emission with respect to the
$V_{\rm LSR}$ of the cloud ($|V_{\rm CO}-V_{\rm LSR}| <$ 8 km s$^{-1}$), and the lack of
a significant amount of redshifted gas, are more compatible with the
molecular outflow being close to the plane of the sky. The absence of
a coherent velocity gradient of the water masers along their linear
structure is also consistent with this geometry.

\subsection{The outflow traced by water masers}
The water masers are distributed in the direction of the extended
infrared nebula, approximately NE-SW.  This is the same
orientation as the blueshifted CO lobe with respect to the central
source, and suggests that the masers are tracing the base of the
outflow.  Note, however, that most of the maser components are
redshifted with respect to the cloud velocity, and yet their
location primarily to the south and west of the infrared source 
means that they coincide with the blueshifted CO lobe.  This 
apparent kinematic discrepancy could be explained if the masers 
are tracing the background walls of a cavity evacuated in the cloud
by a molecular outflow close to the plane of the sky. 
 The existence of such a cavity was suggested by
\citet{Hir97} form their C$^{18}$O map. In the proposed geometry of an outflow close to the
plane of the sky, the background walls could show redshifted
motions, assuming a finite opening angle. The fact that no corresponding blueshifted masers are shown
tracing the foreground walls could be due to different physical
conditions of the gas, with denser gas in the background (that
probably corresponds to the region that goes deeper into the molecular
cloud), and would
produce a stronger interaction with the outflow in that area.
%In this scenario, the fact that we are observing mostly
%redshifted masers could be explained by the existence of denser material to the NE of the infrared 
%source. \citet{Hir97} shows a C$^{18}$O map around B1-IRS with a region
%of higher column density toward the east of the CO blueshifted
%outflow and suggested that there is a cavity of denser gas at the CO
%outflow lobe. This higher density could produce a more violent interaction between 
%winds and the ambient medium, which could favor the pumping of intense maser
%emission in that direction. Moreover, possible blueshifted masers 
%would be less intense, and may have not been detected due to an insufficient
%sensitivity of the measurements. If this is the case, redshifted CO velocities %could be as low as to prevent the corresponding lobe being detected, as 
%suggested by \citet{Hir97}. 
%Another possibility is that the maser structure could be associated to
%the CO blueshifted lobe and delineating the walls of the cavity
%opened by a jet structure. The geometry near the
%plane of the sky could explain the superposition of three blueshifted masers spots
%with the redshifted masers, and the existence of a cavity of denser
%material would support the stronger interaction with redshifted
%material at the back side of the CO lobe. 
Proper motion measurements of the B1-IRS water masers will help to clarify 
wether our proposed geometry, with the outflow close to the plane
of the sky and evacuating a cavity in the cloud, is correct.

\subsection {The interaction between the molecular outflow and its
surrounding environment}

The CCS clumps show a clear velocity gradient, with less
redshifted velocities towards the central source (see
Fig.~\ref{CCSvelocity}). If we assume that the ambient cloud velocity 
is 6.3 km s$^{-1}$ \citep{Hir97}, these clumps would be redshifted with
respect to the cloud velocity. 
This velocity pattern cannot be explained purely by foreground clumps with 
infalling motions, for which we would expect more redshifted velocities closer
to the source. Nevertheless, if we consider that the CCS clumps may be
interacting with the outflowing material, the infalling clumps could be slowed down closer
to the central source due to this interaction, explaining in this way the
less redshifted velocities at these positions.  

An alternative scenario to explain the CCS kinematics is outflowing
clumps driven by a wind. 
If this wind exhibits a Hubble-flow type
velocity, that increases with distance from the driving source (as it is
observed in other outflows from class 0 protostars, \citealt{Cha01}), we could
explain the velocity pattern by background clumps interacting with, and being
accelerated by the outflowing material. In order to have redshifted
CCS emission at both to the NE and SW from the central source, 
the outflow must lie near the plane of the sky, and the CCS clumps 
should be associated with the back side of
the outflow, both for the southwestern blueshifted CO lobe, and the (as yet)
undetected northeastern redshifted CO lobe. 

A way to distinguish between these two models (foreground 
infalling clumps decelerated by the outflow vs. background outflowing 
accelerated clumps), is to ascertain whether the motions observed in
the CCS clumps can be gravitationally bound.

There are two main components in the observed 
motions: the internal velocity gradients within the clumps, and the
bulk velocity of the whole clump with respect to the central source.
For the first component, the mass M needed to  gravitationally bind a
cloud of size $R$ with velocity gradient $dV/dl$ is $M = V^{2} R G^{-1}$, with $V=(dV/dl)R$. 
This means that the masses needed to bind the internal gradient within
each clump are 1.3, 0.3, and 0.4 $M_{\sun}$ for clumps SW, NW, and NE,
respectively. 

The masses derived for the clumps (Table~\ref{tbl-physical}) are of
the same order of those necessary for the clumps to be bound, although
we must consider this result carefully due to the very large
uncertainties involved in these calculations. 
 
On the other hand, if we consider an
average velocity of each CCS clump with respect to that of the ambient cloud,
and the distance from the infrared source to the center of each clump,
the masses needed to bind the motions respectively for SW, NW, and NE CCS clumps as a whole 
are 6, 0.4, and 6
$M_{\sun}$ respectively. The mass of gas contained in a region of
radius equal to the distance to the central source would be  
13-30 $M_{\sun}$ for the SW clump, 5-11 $M_{\sun}$ for the NW clump, and
51-120 $M_{\sun}$ for the NE clump, 
depending on the values of $N_{H_{2}}$ obtained from either 
averaged values for NH$_{3}$ or CCS. 

Therefore, we see that the observed motions in the CCS clumps 
can be gravitationally bound, a fact that does not allow us to discard
either of the alternative models.
Regardless of the real geometry and dynamics of the region, it seems
that there is an important interaction of the ambient gas with the molecular
outflow.

We must point out that the observed CCS clumps may not really be
physical entities. These clumps are likely to represent regions of
enhanced abundance of CCS \citep{Suz92, Oha99}. This chemical gradient
of CCS is a result of variations in the local conditions of the cloud
\citep{Lai03}. However, our mass estimates to check if the motions are
gravitationally bound
are still valid even if the enhanced CCS emission are not physical
clumps, given that we are calculating the mass of the region
of gas enclosed by the CCS emission. The largest source of
uncertainty is the value of the molecular abundance, which is a common
problem when deriving column densities using other molecular tracers.
In our calculations we consider a fractional abundance of CCS with
respect to the H$_{2}$ of 0.9$\times$10$^{-10}$, that is the mean value
reported by \citet{Lai00} for this region.

\subsection{Chemical evolution: CCS vs. NH$_3$}

Single-dish observations show that the CCS is more abundant in
starless, cold, quiescent dark cores (such as \objectname{L1521B},
\objectname{L1498}, \objectname{TMC-1C} or \objectname{L1544};
\citealt{Suz92}) while ammonia emission tends to be intense in star-forming
regions. These results suggested that CCS is a tracer of early time 
chemistry (t $\leq$ 10$^{5}$ yr), while ammonia is a tracer of more evolved gas 
\citep{Hir92,Vel95,Ber97}.

The abundance ratio
$N_{\rm NH_{3}}/N_{\rm CCS}$ has therefore been proposed by \citet{Suz92}
as an indicator of the evolutionary stage of molecular clouds.
Our single dish data  provides
column densities of $N_{\rm NH_{3}} \simeq (6\pm 4)\times 10^{14}$ and
$N_{\rm CCS} \simeq (1.70\pm 0.24)\times 10^{13}$ cm$^{-2}$. Thus, the
ratio $N_{\rm NH_{3}}/N_{\rm CCS}$ is $\simeq$35, which is
significantly lower than the value $\geq 100$ found by \citet{Suz92}
in most of the star forming regions in their survey.
This relative lack of ammonia, or enhanced CCS, may be an
indication of the early evolutionary stage of B1-IRS. However this result
could depend on the particular chemistry of each region, 
and we must be careful in this interpretation.

In previous single-dish works, a spatial anticorrelation between CCS
and ammonia was found, both by comparing detections of those molecules
in surveys of sources \citep{Suz92}, and by mapping their emission in
the same region \citep{Lai03}. It is interesting to check whether this
 anticorrelation still stands when we map a source at high
 resolution. The overlay of our CCS and ammonia maps
 (Fig. \ref{CCS-NH3}) shows a clear anticorrelation
between them throughout the B1-IRS region, except $\simeq30''$ to the
north of the ammonia
maximum (which coincides with the NE CCS clump), where there is some
superposition. This is the first time that this anticorrelation is observed 
at such a high-resolution ($\simeq 5''$ = 1750 AU) in a star-forming
region, and illustrates that the chemistry is not only time-dependent,
but also exhibits small-scale spatial structure that can potentially
seriously affect the interpretation of derived physical and dynamical
properties if it is not properly taken into account.

The chemical gradients traced by these molecules may be the result of 
changing the local conditions. At this moment, most of the theoretical
chemical models considering the age and distance to the cloud center,
include a large number of physical parameters
 to study the variation of the column density, 
the distribution, and the fractional abundance of molecules like 
CCS and NH$_{3}$ among other species.
Some of these models include the strength of magnetic fields, 
initial chemical composition of the cloud, the probability of
molecules sticking to dust grains, 
cosmic-ray ionization
rate, and cloud mass \citep{She03}, UV photodisociation,
cosmic-ray induced photoionization and photodisociation reactions, CO
depletion onto grains, and desorption by cosmic-rays \citep{Nej99}, or
core geometry and view angles \citep{Aik03}. Until now, these chemical
studies focused on starless cores and did not consider the 
effects of a central star, probably because most of the CCS studies 
have been made toward starless cores \citep{Kui96,Oha99,Lai03,Shi04}. 

The way in which all of these factors and the activity of protostars 
(in the case of star forming regions) can modify the classical ion-molecule or 
radical/neutral pathways of production of molecules like CCS \citep{Smi88,Suz88,Pet96,Sca96} needs to be studied in theoretical
works to give a better interpretation of our results.
The development of models that implement the onset of energetic activity
associated with the formation of a star and how this affects
the CCS and NH$_{3}$ production will be the key to ascertain
 the evolutionary stage of B1-IRS.
Models treating consistently both CCS and NH$_{3}$ and considering the
effect of young stars could provide more reliable values of the
relative abundances and column densities of both molecules as a function of the
distance from the source. On the other hand, our data on B1-IRS could be
used as a reference work to test the predictions of those calculations
in detail. Future comparison  with other regions observed with
high angular resolution will be vital for understanding  the
precise meaning of the ratio between CCS and NH$_{3}$ abundances in
star-forming regions, in terms of chemical evolution.

Moreover, detailed chemistry studies in molecular clouds are an
important science driver for the development of new interferometers (e.g. SMA
or ALMA, see \citealt{Van98,Phi03}). Our anticorrelation result at
small scale proves that chemistry studies
is indeed a promising line of research for these new telescopes.

The clumpy distribution of CCS emission detected in B1-IRS, has already
been observed by other authors in several molecular clouds such as 
\objectname{B335} \citep{Vel95}, \objectname{L1498} \citep{Kui96}, and \objectname{TMC-1} \citep{Lan95}.  
In the case of B1-IRS, this clumpy distribution was first detected 
by \citet{Lai00} in the CCS J$_{N}$= 3$_{2}$-2$_{1}$ transition at
33.8 GHz with the BIMA interferometer with an angular resolution of
$\simeq$ 30$''$. In particular, there seems to be a spatial coincidence
between BIMA clumps F and B and our VLA clumps SW and NE respectively.
Moreover our VLA observations reveal a lack of CCS emission at the
position of B1-IRS. This
absence of emission has been suggested to be due to a lower abundance of
the molecule, rather than to a density decrease, 
since this molecule needs high density to be excited and the density is
expected to be higher towards the center \citep{Vel95}. 
The clumpy distribution could be due to an episodic infall 
(as suggested by \citealt{Vel95} for B335). 

%Another possibility that it is worth to be mentioned is that perhaps the CCS could be enhanced 
%in shocked regions. This is a possibility that appears to be remote and
%apparently argues against the typical association to the CCS with cold and quiescent
%medium, and need to be carefully tested by theoretical calculations.
%There are precedents in the literature that indicate that the
%molecular outflows affect the chemistry of the surrounding
%medium. \citet{Tor92,Tor93} as result of interferometric observations
%in a outflowing star forming region, suggested the possibility
%that the NH${3}$ emission could be enhanced in shocked regions near a Herbig
%Haro object, and this prediction was confirmed soon after
%with theoretical models by \citet{Tay96}. Another good example of how
%an outflow may affects the chemistry of its surrounding medium is the interferometric observations
%in CS reported by \citet{Arc04} that show this molecule delineating the outflow
%cavity walls. In this case theoretical models firstly predicted an
%abundance enhancement of that molecule at the interface between outflows and molecular
%cores due to shocks \citep{Vit02}.  

Another possibility that it is worth considering is that CCS might
be locally enhanced in shocked regions. 
In fact, the observed CCS clumps show kinematical signs
of interaction with the outflow, which suggests that 
the CCS emission could be originated from gas affected by the
impact of shocks. The outflow could squeeze low
density gas to the higher densities ($\simeq$ 10$^{5}$ cm$^{-3}$) needed to produce the CCS, and trigger
the formation of the molecule. Such low density gas exists around dense clouds like B1,
and the outflow associated to B1-IRS could interact with fresh material around the
dense core. 
Although there is currently no chemical model that includes this kind
of phenomena related to the star formation, to support the suggested enhancement,
 observations of other molecules have suggested that their abundance is enhanced in the
environment of jets and outflows, by shock-induced chemistry, as in
the case of CS
\citep{Arc04}, NH$_3$ \citep{Tor92, Tor93}, or HCO$^+$ \citep{Gir00}. Theoretical chemistry models 
support those findings \citep{Tay96, Vit02, Wol93}. In particular, for HCO$^+$ it has been proposed that
in some cases its abundance may be increased by a fast outflow impinging
on fragments of dense gas \citep{Raw00}, and
this type of interaction between outflow and dense gas is also suggested by
our CCS data, although the chemical reactions involved will certainly
be different. 

If this abundance enhancement takes place in short timescales
($\simeq$ 10$^{5}$ yr), one could expect to find clumps of the order
of 5$''$ at the distance of B1 (350 pc), considering a typical sound
speed of $\simeq$ 0.1 km s$^{-1}$. Therefore, new interferometric
observations of CCS in other star-forming regions would be necessary
to study these processes.

%this kind of studies
%available for CCS, as well as more high resolution observations of
%this molecule in star forming regions.

%The CCS molecule has usually been found to be associated with quiescent 
%dark clouds, while here we suggest a possible association with a molecular
%outflow. If these results are confirmed with observations at
%high spatial resolution in other star-forming regions, 
%they may provide an independent measure of the
%evolutionary state of a cloud, where CCS is associated with the quiescent
%medium in prestellar regions, and with outflowing material around young
%stellar objects. 

The association of the central source with water maser emission  (which
typically lasts for $\simeq$ 10$^{5}$ yr in star forming regions;
\citealt{For89}) and the 
existence of CCS, a molecule characteristic of early time chemistry
\citep{Hir92,Suz92} confirm the early stage of evolution of B1-IRS. The
presence of CCS can be used as a marker of extreme youth among Class 0
protostars (de Gregorio-Monsalvo 2005, in preparation).
Furthermore, this combination of spatial, 
kinematic, and dynamical information from the CCS and ammonia may be
used as a powerful tool to study the processes taking place in the
vicinity of very young Class 0 protostars and not just in quiescent
cores.
 
Clearly, further interferometric CCS
observations of the environment
of other young stellar objects as well as  theoretical chemistry
models will be useful to determine whether CCS
emission tends to selectively trace regions of interaction of the
outflow with its surrounding medium. These studies combined with
high-resolution data of NH$_3$ and other molecular tracers
will also help to determine the spacial distribution and dinamical
evolution of these species.

\section{Conclusions}

We have presented high resolution observations of ammonia, CCS, and water
 masers toward the Class 0 object B1-IRS (IRAS 03301+3057). 
Our main conclusions are as follows:

\begin{itemize}
\item There is a 2MASS infrared point source located $\sim 6 \arcsec$
  NE from the nominal position of the IRAS source, and at the NE 
  tip of a blueshifted CO outflow lobe.  
  Fainter infrared emission extending towards the SW 
  from the point source is also detected.
\item There is a group of water masers associated with the infrared point
  source. Their elongated distribution (in the
  NE-SW direction), and their
  unbound motions suggest that these masers are tracing the base of
  the outflow.
\item The infrared and water maser data suggest that the infrared point
  source traces the position of the powering source of the mass-loss 
  phenomena in the region, and that mass is ejected along the NE-SW direction.
\item We detect three clumps of CCS emission surrounding the
  infrared source, all redshifted with
  respect to the systemic cloud velocity. They show clear velocity gradients,
  with less redshifted gas towards the central source. We interpret
  these gradients in terms of clumps that are strongly interacting
  with a molecular outflow that lies almost in the plane of the sky.
\item Ammonia emission is weak and extended. It shows a spatial
  anticorrelation with CCS. Although this kind of anticorrelation is
  known to be present in other star-forming regions, this is the first
  time that it has been observed with such a high-angular resolution 
  ($\simeq 5''$).
\item The abundance ratio between ammonia and CCS is significantly
  lower than that found in other star-forming regions. Given the fast
  disappearance of CCS in star-forming regions, this low ratio 
  might be an indication of  B1-IRS being an extremely young stellar
  object ($\leq$ 10$^5$ yr).
\item We suggest the possibility that CCS abundance is enhanced via
  shock-induced chemistry. However, theoretical calculations will be 
  needed to confirm this hypothesis.

 \end{itemize}

\acknowledgments

GA, JFG, IdG, and JMT acknowledge support from the Ministry of
Science and Technology (MCYT) grant (European Fund of Regional
Development, FEDER) AYA2002-00376 (Spain). 
GA acknowledges support from Junta de Andaluc\'{\i}a. 
IdG acknowledges the support of a Calvo Rod\'es
Fellowship from the Instituto Nacional de T\'ecnica Aeroespacial (INTA). She
is also thankful to NRAO for their financial support and to the AOC staff
for their hospitality during her stay as a NRAO summer student.
We are thankful to Luis Felipe Rodr\'{\i}guez for his useful comments
on the manuscript.
This paper is partly based on observations taken during 
``host-country'' allocated time at Robledo de Chavela; this time is
managed by the Laboratorio de Astrof\'{\i}sica Espacial y F\'{\i}sica
Fundamental (LAEFF) of INTA, under agreement with National
Aeronautics and Space Administration/Ingenier\'{\i}a y Servicios
Aeroespaciales (NASA/INSA). It also makes use of data products from
the Two Micron All Sky Survey (2MASS), which is a joint project of the
University of Massachusetts and the Infrared Processing and Analysis
Center/California Institute of Technology, funded by NASA and the
National Science Fundation.

\clearpage

\epsscale{0.35}

\begin{deluxetable}{lllcr}
\tabletypesize{\scriptsize}
\tablecaption{Water maser components on 2003 April 04\label{tbl-april}}
\tablewidth{0pt}
\tablehead{
\colhead{Right Ascension\tablenotemark{a}}&
\colhead{Declination\tablenotemark{a}} &
\colhead{Position \tablenotemark{b,c}}&
\colhead{Flux density \tablenotemark{b}}&
\colhead{$V_{\rm LSR}$\tablenotemark{d}}\\
\colhead{(J2000)}  & 
\colhead{(J2000)}&
\colhead{uncertainty ($\arcsec$)}& 
\colhead{(mJy)}& 
\colhead{(km s$^{-1}$)} 
}
\startdata
03 33 16.62    &31 07 54.6    &0.3   &4.3$\pm$1.6   &9.1   \\
03 33 16.629   &31 07 54.89   &0.17  &7.3$\pm$1.7   &3.8   \\
03 33 16.63    &31 07 54.3    &0.3   &4.9$\pm$1.7   &19.0  \\
03 33 16.63    &31 07 54.6    &0.3   &3.9$\pm$1.7   &22.3  \\
03 33 16.636   &31 07 54.58   &0.17  &7.1$\pm$1.7   &10.4  \\
03 33 16.64    &31 07 54.3    &0.4   &3.2$\pm$1.7   &22.9  \\
03 33 16.641   &31 07 54.88   &0.13  &9.7$\pm$1.6   &14.4  \\
03 33 16.643   &31 07 54.98   &0.13  &9.3$\pm$1.6   &11.7  \\
03 33 16.648   &31 07 55.06   &0.09  &13.1$\pm$1.6  &15.0  \\
03 33 16.650   &31 07 54.89   &0.14  &8.6$\pm$1.6   &18.3  \\
03 33 16.65    &31 07 54.9    &0.3   &4.7$\pm$1.7   &16.4  \\
03 33 16.656   &31 07 54.88   &0.09  &14.7$\pm$1.7  &12.4  \\
03 33 16.657   &31 07 55.01   &0.08  &15.4$\pm$1.6  &17.7  \\
03 33 16.659   &31 07 55.02   &0.16  &7.8$\pm$1.7   &3.2   \\
03 33 16.659   &31 07 55.17   &0.19  &6.4$\pm$1.6   &11.1  \\
03 33 16.665   &31 07 55.09   &0.17  &6.9$\pm$1.6   &9.8   \\
03 33 16.666   &31 07 55.15   &0.12  &10.6$\pm$1.6  &13.7  \\
03 33 16.670   &31 07 54.99   &0.10  &12.3$\pm$1.7  &13.1  \\
03 33 16.67    &31 07 55.2    &0.3   &3.6$\pm$1.7   &2.5   \\
03 33 16.68    &31 07 56.3    &0.3   &4.0$\pm$1.6   &23.6  \\
03 33 16.684   &31 07 54.92   &0.12  &7.1$\pm$1.7   &17.0  \\
03 33 16.684   &31 07 55.02   &0.13  &9.2$\pm$1.6   &15.7  \\
03 33 16.70    &31 07 55.1    &0.3   &4.1$\pm$1.6   &19.7  \\
\enddata
\tablenotetext{a}{Units of right ascension are hours, minutes, and
  seconds. Units of declination are degrees, arcminutes, and arcseconds }
\tablenotetext{b}{Uncertainties are $2\sigma$}
\tablenotetext{c}{Relative position uncertainties with respect to the
  phase center. The absolute position error of the phase center is $\sim 0\farcs 28$ }
\tablenotetext{d}{Velocity of maser emission}
\end{deluxetable}

\begin{deluxetable}{lllcr}
\tabletypesize{\scriptsize}
\tablecaption{Water maser components on 1998 October 24\label{tbl-oct}}
\tablewidth{0pt}
\tablehead{
\colhead{Right Ascension\tablenotemark{a}}&
\colhead{Declination\tablenotemark{a}} &
\colhead{Position \tablenotemark{b,c}}&
\colhead{Flux density \tablenotemark{b}}&
\colhead{$V_{\rm LSR}$\tablenotemark{d}}\\

\colhead{(J2000)}  & 
\colhead{(J2000)}&
\colhead{uncertainty ($\arcsec$)}& 
\colhead{(mJy)}& 
\colhead{(km s$^{-1}$)} 
}

\startdata
03 33 16.6493   &31 07 55.049  &0.024  &440$\pm$80   &15.1  \\
03 33 16.6498   &31 07 55.039  &0.020  &570$\pm$90   &14.8  \\ 
03 33 16.6504   &31 07 55.080  &0.017  &750$\pm$100   &14.5  \\ 
03 33 16.6509   &31 07 55.061  &0.019  &560$\pm$80   &14.2  \\ 
03 33 16.651    &31 07 55.10   &0.03   &610$\pm$90   &13.8  \\ 
03 33 16.6538   &31 07 55.062  &0.009  &2300$\pm$300   &16.1  \\
03 33 16.654    &31 07 55.04   &0.03   &360$\pm$70   &13.5  \\ 
03 33 16.6591   &31 07 55.067  &\nodata\tablenotemark{e} &2800$\pm$300   &15.8  \\
03 33 16.6593   &31 07 55.069  &0.011  &970$\pm$130   &15.5  \\
03 33 16.661    &31 07 55.06   &0.15   &1050$\pm$140   &16.5  \\
03 33 16.667    &31 07 55.11   &0.07   &130$\pm$60   &16.8  \\
\enddata
\tablenotetext{a}{Units of right ascension are hours, minutes, and
  seconds. Units of declination are degrees, arcminutes, and arcseconds }
\tablenotetext{b}{Uncertainties are $2\sigma$}
\tablenotetext{c}{Relative position uncertainties with respect to the
  reference feature used for self-calibration}
\tablenotetext{d}{Velocity of maser emission}
\tablenotetext{e}{Reference feature. Absolute position error $\sim 0\farcs 021$ }
\end{deluxetable}

\begin{deluxetable}{lllcr}
\tabletypesize{\scriptsize}
\tablecaption{Water maser components on 1999 February 25\label{tbl-feb}}
\tablewidth{0pt}
\tablehead{
\colhead{Right Ascension\tablenotemark{a}}&
\colhead{Declination\tablenotemark{a}} &
\colhead{Position \tablenotemark{b,c}}&
\colhead{Flux density \tablenotemark{b}}&
\colhead{$V_{\rm LSR}$\tablenotemark{d}}\\

\colhead{(J2000)}  & 
\colhead{(J2000)}&
\colhead{uncertainty ($\arcsec$)}& 
\colhead{(mJy)}& 
\colhead{(km s$^{-1}$)} 
}
\startdata
03 33 16.591   &31 07 54.91   &0.16       &137$\pm$23   &19.1 \\
03 33 16.628   &31 07 54.69   &0.14       &118$\pm$20   &12.8 \\
03 33 16.636   &31 07 55.11   &0.25       &57$\pm$27   &20.1 \\
03 33 16.641   &31 07 54.95   &0.04       &352$\pm$24   &15.1 \\
03 33 16.641   &31 07 54.96   &0.04       &362$\pm$21   &13.2 \\
03 33 16.646   &31 07 55.00   &0.05       &36$\pm$30   &17.4 \\
03 33 16.646   &31 07 55.07   &0.13       &142$\pm$22   &18.1 \\
03 33 16.6465  &31 07 55.017  &0.014      &1180$\pm$50   &17.1 \\
03 33 16.647   &31 07 55.02   &0.16       &152$\pm$23   &18.4 \\
03 33 16.6470  &31 07 55.006 &\nodata\tablenotemark{e} &6500$\pm$230   &15.8 \\
03 33 16.6481  &31 07 55.021  &0.003      &6010$\pm$220   &16.1 \\
03 33 16.6481  &31 07 55.026  &0.007      &2090$\pm$80   &15.5 \\
03 33 16.6483  &31 07 55.013  &0.006      &2810$\pm$100   &16.5 \\
03 33 16.649   &31 07 54.99   &0.04       &410$\pm$30   &13.5 \\
03 33 16.649   &31 07 55.03   &0.06       &236$\pm$23   &13.8 \\
03 33 16.6493  &31 07 55.017  &0.008      &1810$\pm$70   &16.8 \\
03 33 16.65    &31 07 55.1    &0.3        &45$\pm$24   &19.8 \\
03 33 16.653   &31 07 54.95   &0.17       &98$\pm$23   &17.8 \\
03 33 16.660   &31 07 55.04   &0.10       &169$\pm$23   &14.2 \\
03 33 16.66    &31 07 55.3    &0.4        &27$\pm$20   &11.2 \\
03 33 16.663   &31 07 55.08   &0.13       &17$\pm$30   &19.4 \\
03 33 16.663   &31 07 55.20   &0.14       &111$\pm$22   &18.8 \\
03 33 16.666   &31 07 54.87   &0.11       &108$\pm$21   &14.8 \\
03 33 16.684   &31 07 54.95   &0.14       &104$\pm$20   &14.5 \\
\enddata
\tablenotetext{a}{Units of right ascension are hours, minutes, and
  seconds. Units of declination are degrees, arcminutes, and arcseconds }
\tablenotetext{b}{Uncertainties are $2\sigma$}
\tablenotetext{c}{Relative position uncertainties with respect to the
  reference feature used for self-calibration}
\tablenotetext{d}{Velocity of maser emission}
\tablenotetext{e}{Reference feature. Absolute position error $\sim 0\farcs 12$}
\end{deluxetable}

\begin{deluxetable}{llccccrc}
\tabletypesize{\scriptsize}
\tablecaption{Physical parameters of the clumps observed in NH$_3$ and CCS\tablenotemark{*} \label{tbl-physical}}
\tablewidth{0pt}
\tablehead{
\colhead{Molecule}&
\colhead{Clump}&
\colhead{I$_{\nu}$\tablenotemark{a}}&
\colhead{$\int{I_\nu dv}$\tablenotemark{a}} &
\colhead{$N_{\rm mol}$\tablenotemark{b}}& 
\colhead{$N_{\rm H_{2}}$\tablenotemark{c}}& 
\colhead{Sizes\tablenotemark{d}} &
\colhead{$M_c$\tablenotemark{e}}\\   
 & &
\colhead{(mJy beam$^{-1}$)}&
\colhead{(mJy beam$^{-1}$ km s$^{-1}$)}& 
\colhead{($10^{13}$ cm$^{-2}$)}&
\colhead{($10^{23}$ cm$^{-2}$)}&
\colhead{(AU$\times$ AU)}&
\colhead{($M_\sun$)}

}

\startdata
NH$_{3}$ & Mean\tablenotemark{f}   &170$\pm$30 &210$\pm$30  &140$\pm$40 &1.4$\pm$0.4  &10700$\times$4600 &2.1$\pm$0.6 \\  
CCS &SW1  &14.2$\pm$2.8   &7.5$\pm$1.4          &3.7$\pm$1.7 &4.1$\pm$1.9 &2900$\times$2900  &1.1$\pm$0.5\\ 
CCS &SW2  &10.5$\pm$2.8   &6.0$\pm$1.4          &2.9$\pm$1.5 &3.2$\pm$1.7 &5000$\times$2900  &1.4$\pm$0.7\\
CCS &NW   &4.7$\pm$2.6    &3.4$\pm$1.3          &1.7$\pm$1.1 &1.9$\pm$1.2 &7000$\times$2100  &0.8$\pm$0.5\\ 
CCS &NE   &9.3$\pm$3.9    &7.3$\pm$2.0          &3.6$\pm$1.7 &4.0$\pm$1.9 &12600$\times$2900 &4.4$\pm$2.0\\ 

\enddata

\tablenotetext{*}{All the uncertainties in this table are 2$\sigma$.} 
\tablenotetext{a}{Intensity and integrated intensity at the position
  of the emission peak for each clump (main  hyperfine component only for ammonia).}
\tablenotetext{b}{Column density obtained from $N_{\rm {mol}} = \frac
  {8\alpha \pi \nu^{3}}{c^{3} g_{j} A_{ji}} \frac{Q(T_{\rm{rot}}) \int {I_\nu
      dv}}{[B_\nu(T_{\rm {ex}}) - B_\nu(T_{\rm
      {bg}})]}\frac{\exp({E_{j}/kT_{\rm{rot}}})}{\exp(h\nu/kT_{ex})-1}$
  (optically thin approximation), 
   where $\alpha$ is a factor equal to 1 for CCS, and equal to
$2[1+\exp(h\nu/kT_{\rm ex})]$ for NH$_{3}$, $\nu$ is the frequency of the transition, $Q$
is the partition function, $\int {I_\nu dv}$ is the integrated
intensity (referred only to the main line in the case of the ammonia),
E$_{j}$ is the energy
of the upper state in the case of the CCS (1.61 K; \citealt{Wol97}) and the energy of the
rotational level whose inversion transition is observed in
the case of the ammonia (23.4 K; \citealt{Ho83}), $T_{\rm rot}$ is the rotational temperature,
g$_{j}$ is the statistical weight of the upper rotational level in the
case of the CCS, and of the upper sublevel involved in the inversion
transition in the case of the ammonia.
$A_{ji}$ is the Einstein coefficient for the overall transition (1.67 and 4.33 $\times$ 10$^{-7}$s$^{-1}$ 
for the NH$_3$(1,1) and CCS(2$_1$-1$_0$) transitions, respectively;
\citealt{Ho83,Wol97}), and $T_{\rm ex}$
is the excitation temperature. For the CCS molecule we adopted a
  $T_{\rm rot}$ = $T_{\rm ex}$ = 5 K  \citep 
{Suz92}. For the NH$_{3}$ calculations,
we have assumed a $T_{\rm rot}$ = 12 K derived by  \cite
 {Bac84} from single-dish observations. From our single dish ammonia spectrum (see Fig. \ref{Single_dish}),
we derived an optical depth of $\tau = 0.9\pm 0.5$, which provides a
$T_{\rm ex} = 8.8\pm 2.4$ K. The final $N_{\rm NH_{3}}$ was obtained by
  multiplying the
  optically 
  thin solution by the factor
  $(\tau/(1-e^{-\tau})$).} 
\tablenotetext{c}{Hydrogen column density, obtained assuming a fractional abundance
  with respect to H$_2$ of 10$^{-8}$ for NH$_{3}$ \citep{Her73} and 0.9
  10$^{-10}$ for CCS \citep{Lai00}.}  
\tablenotetext{d}{FWHM of the clumps. It is an averaged value in the
  case of the ammonia.}
\tablenotetext{e}{Clump mass, derived from the $N_{H_{2}}$ and the half-power
  area. It is an averaged value for ammonia clumps.}
\tablenotetext{f}{Average value for the NH$_3$ clumps.}

\end{deluxetable}

\end{document}